# A NOVEL APPROACH FOR MOBILITY MANAGEMENT IN LTE FEMTOCELLS

[1]Pantha Ghosal, [2]Shouman Barua, [3]Ramprasad Subramanian, [4]Shiqi Xing and
[5]Kumbesan Sandrasegaran

[1,2,3,4,5,6]Centre for Real-time Information Networks
School of Computing and Communications, Faculty of Engineering and Information Technology, University of Technology Sydney, Sydney, Australia

pantha.ghosal@uts.edu.au
shouman.barua@uts.edu.au
ramprasad.subramanian@uts.edu.au
shiqi.xing-1@student.uts.edu.au
kumbesan.sandrasegaran@uts.edu.au

## ABSTRACT

*LTE is an emerging wireless data communication technology to provide broadband ubiquitous Internet access. Femtocells are included in 3GPP since Release 8 to enhance the indoor network coverage and capacity. The main challenge of mobility management in hierarchical LTE structure is to guarantee efficient handover to or from/to/between Femtocells. This paper focuses, on different types of Handover and comparison performance between different decision algorithms. Furthermore, a speed based Handover algorithm for macro-femto scenario is proposed with simulation results*

## KEYWORDS

*Femtocell Access Point (FAP), Handover Hysteresis Margin (HMM), Reference Signal Received Power (RSRP), Reference Signal Received Quality (RSRQ), Signal to Interference Plus Noise Ratio (SINR, Evolved NodeB (eNB), User equipment (UE).*

## 1. INTRODUCTION

In the next generation wireless communication systems, the primary challenge is to improve the indoor coverage, capacity enhancement as well as to provide users the mobile services with high data rates in a cost effective way. Performance of mobile system can be enhanced by evolving to emerging broadband technologies such as WiMAX [1] and LTE [2] but this may not be able to endure the exponential rise in traffic volume. These advancements in 4G physical layer (PHY) are approaching to the Shannon limit [3] and ensure maximum achievable data rate. So, further enhancement either in the PHY layer or available spectrum will not be adequate to overcome the coverage and capacity challenge. One of the approaches of solving this capacity and coverage related problem includes moving the transmitters and receivers closer to each other. This method loses its ground on economic feasibility because of deploying more base stations (BSs). Thus, small cells generally known as Femtocells with restricted access permission to fewer users compared to macrocell are chosen by the mobile operators as a possible solution to improve the network coverage, especially to the indoor users with ubiquitous high speed connectivity. These Femtocell base stations are referred to as Femto Access Points (FAPs). They have a short-range (10-30m) and require a low power (10-100mW) [5] to provide high-bandwidth wireless communication services in a cost effective way. Femtocells incorporated with the plug and play capabilities work in mobile operator owned licensed spectrum and enable Fixed Mobile Convergence (FMC) [6] by connecting to the core network via broadband communications links (e.g., DSL). Unlike macrocells, FAPs are typically installed and maintained by the end users in an unplanned manner and don't have X2 interface between them for information sharing. Due to this uncoordinated nature femtocell pose challenge on Handover and Radio Resource Management. The rest of the article is organized as follows: Section.2 describes the LTE Network architecture with Femtocells, Section.3 depicts the HO types and open challenges in HO management, Section 4. Describes different HO algorithms and their performance comparison and in Section. 5 proposed HO algorithm with simulation result is described.

## 2. LTE NETWORK ARCHITECTURE

The 3GPP, LTE is a packet-switched with flat architecture and is mainly composed of three parts: the UE, the e-UTRAN, and the packet switched core network (EPC). EPC is responsible for all services provided to UE including voice and data to the user using packet switching technology. e-UTRAN has only one node i.e., the evolved NodeB (eNB) which handles the radio communication between UE and EPC. The physical layer of radio interface supports both time (TDD) and frequency (FDD) division duplexing. On the other hand, Femtocells not only boost indoor coverage [9] and capacity but also improve battery life of UEs since UE doesn't need to communicate with a distant macrocell base stations. Fig. 1(a) shows the basic two tier macro-femto network architecture and Fig. 1(b) shows X2 and S1 interfaces. FAPs which have a less computational capability [3], are connected through DSL (Digital Subscriber Line) in indoor scenarios. The LTE macro system based on flat architecture connects all the eNBs through X2 interface and the RRM procedure is done by eNB.

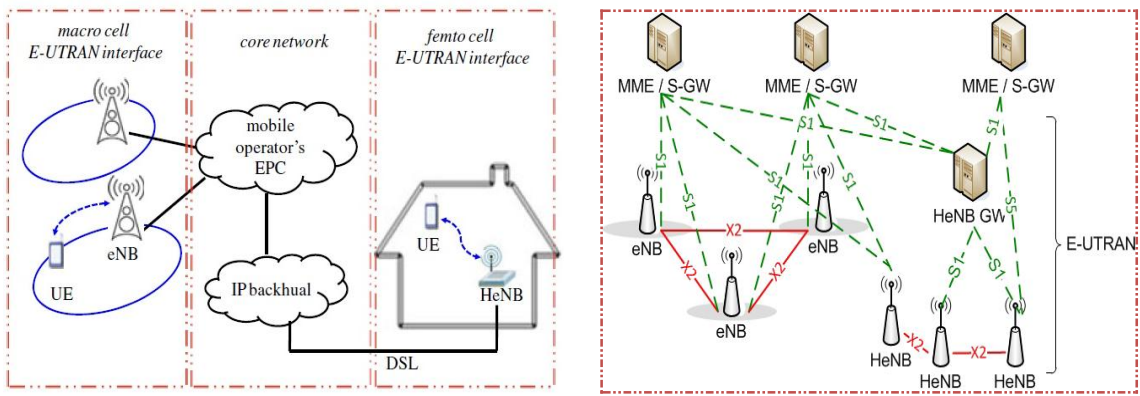

Figure 1:a) Two-tier macro-femto network architecture [3], b) Network Interfaces[9]

Femtocells can to operate in one of three access modes, i.e., closed access mode, open access mode or hybrid access mode [7]. Closed access mode is generally deployed in residential scenarios and a group of registered users called Closed Subscriber Group (CSG) have the permission to access the femtocell. In case of open access mode, any UE can access the femtocell and benefit from its services but when it comes to resource usage, congestion and security, open access is not a suitable choice. In hybrid access mode, a limited number of unknown MUE may access the femtocell while a fixed number of users defined by the owner can access the femtocell ubiquitously but may suffer the risk of security breach [8]. In this article, Closed Access Mode is considered because of security and privacy of the owner.

## 3. Handover to Femtocells and Challenges

In the two tier macro-femto scenario there are three possible handover scenarios as shown in Fig.2. When a UE is moving in femtocell coverage from macrocell coverage the HO that takes place is called Inbound HO and Outbound HO is one where UE gets out from the femtocell coverage to the macrocell coverage. Femtocell user (FUE) moving one FAP coverage area to another FAP coverage area is called Inter-FAP HO. LTE doesn't support X2 interface between eNB and FAP and the large asymmetry in received signal strengths makes inbound and inter-FAP HO more complex. While inbound HO, apart from received signal strength and signal quality access control, interference, user speed and position has to be taken in consideration. On the other hand, since UE moves from femtocell coverage to the macrocell coverage stored in its neighbour list with best received signal strength, the outbound HO procedure is not that complicated. The HO phase for two-tier system can be divided in to six phases[9] : 1)Cell identification 2)access control 3) cell search 4) cell selection/reselection 5)HO decision 6) HO execution. The HO phases are shown in Fig.3. The position of the femtocells are known since they are connected to the network through backhaul and the whitelist of accessible FAPs are stored in user U-SIM[9]. When user comes near to FAP coverage it gets a proximity notification from the network and collect signal information according to eNB prescribed measurement configuration. FAPs are identified through their physical cell id(PCI). The number of PCI is limited, which is only 504. So, in case of unplanned and dense deployment of FAPs, cell selection/resection/search it may become confusing to choose the accessible FAP when there are more than one FAP with the same PCI. Unable to perform that may increase the number of HO failure [9]. As it can be seen from Fig. 3 with the presence of Home eNB gateway (which is used for UE authenticate and access control) two additional steps are taken while HO decision and Execution. The additional steps as illustrated in Fig. 3($\oplus$ sign) increase the delay in connection setup.

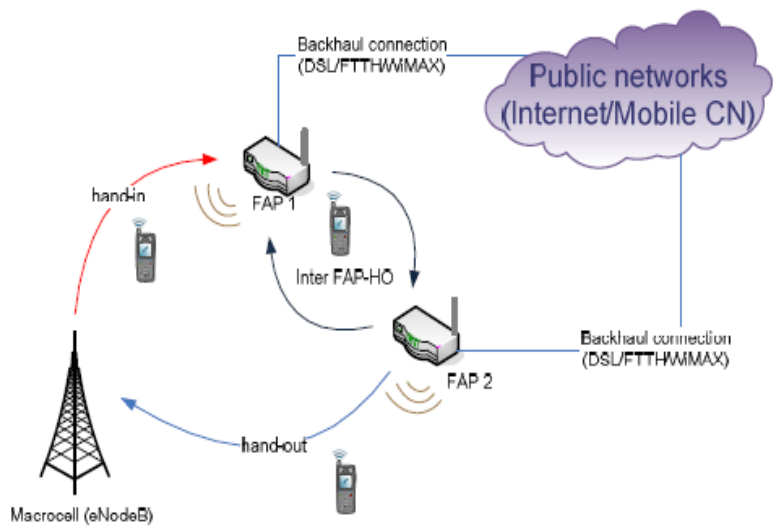

Fig. 2: Handover Scenario in presence of FAP [14]

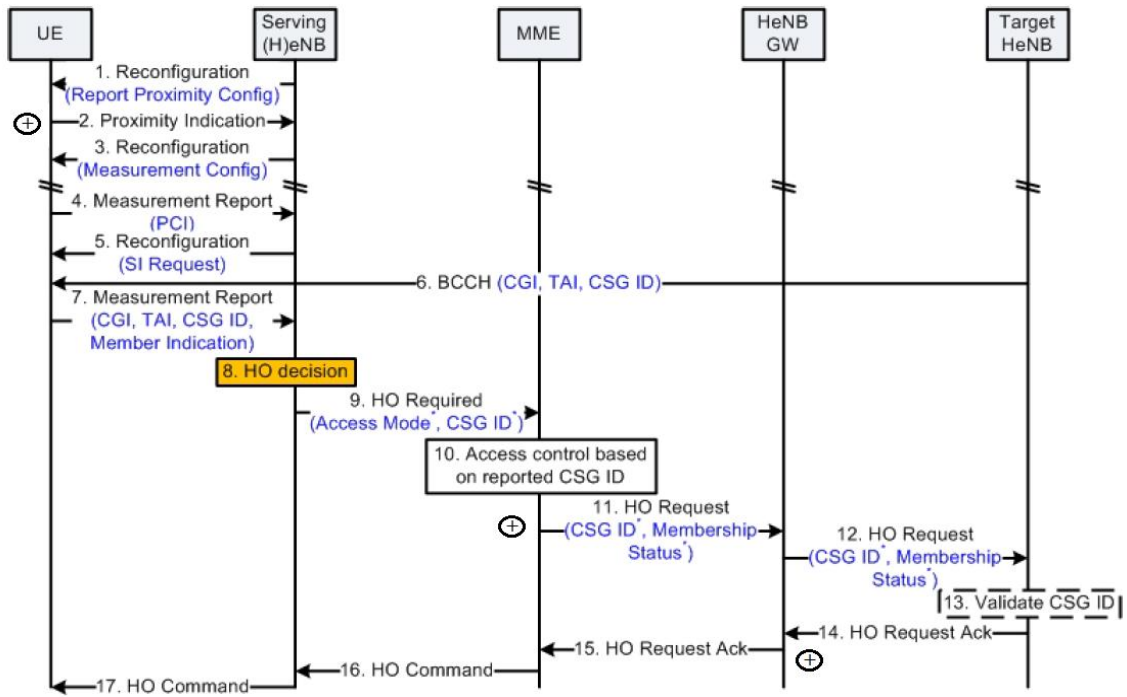

Fig. 3: HO execution signalling procedure for inbound mobility to a FAP [9]

## 4. Handover Decision Algorithms

The Handover decision criteria form macro-femto two tier network is different than the macro cellular network since there is no direct interface like X2 between them. In this paper, different proposed HO algorithms based on different parameters are discussed. The main decision parameter for handover to/from/between FAPs [9] can be divided in in five groups: Received signal strength (RSS/RSRP), b) User speed, c) cost-function based, d) Interference experienced at user end or serving cell, and e) Energy Emission. Since, FAPs have less computational capability and also prone to delay and congestion due to external backhaul (DSL), in this paper we have considered received signal strength, speed and interference for evaluation and evolving proposed HO algorithm.

### 4.1.1 Received Signal Strength Based HO Algorithm:

The handover decision algorithms in this class are based on the Received Signal Strength as shown in Fig. 4. To minimize the HO probability and ping-pong effect the RSS based algoritihms consider a HO Hysteresis Margin (HMM) to compare the received signal strength of the source and target cells.

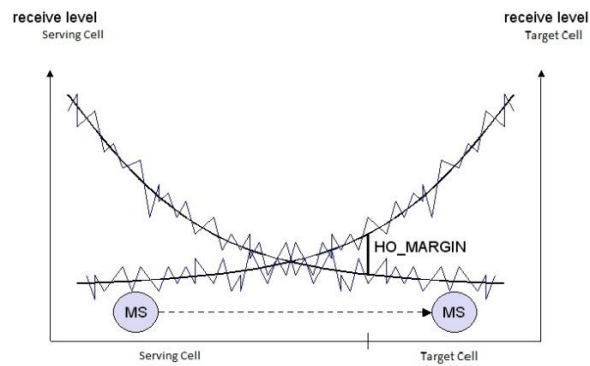

Fig. 4: Conventional Received Signal Strength Based Handover

When an UE enters the coverage area of femtocells it experiences uneven transmission powers from macrocell and femtocell. The proposed algorithm in [10], Compensates the uneven RS power transmission in single macrocell-femtocell scenario by combining RSS of the macrocell and femtocell. This algorithm uses an exponential window function to mitigate the first variation of RSS. The operation can be expressed as follows:

$$\bar{s}_m[k] = w[k] * s_m[k] \quad (1)$$
$$\bar{s}_f[k] = w[k] * s_f[k] \quad (2)$$

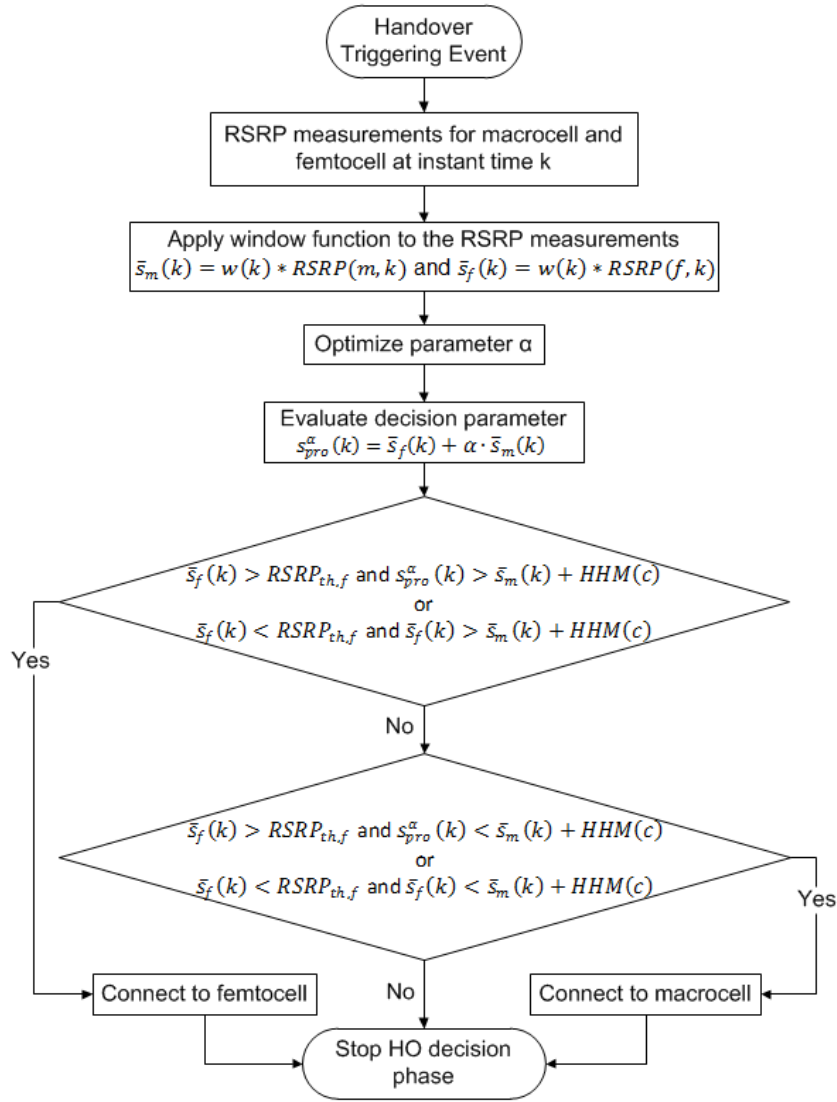

Fig. 5: HO Algorithm based on RSS [10]

Here, *w(k)* denotes the exponential window function and $\bar{s}_m[k]$, $\bar{s}_f[k]$ represents the filtered RSS parameters of the macrocell and the femtocell at time *k* respectively. These filtered signals are then combined into a RSS-based decision parameter as follows:

$$s_{pro}^{\alpha}[k] = \bar{s}[k] + \alpha \bar{s}_m[k] \qquad (3)$$

Where $a \in [0, 1]$ is the combination factor to compensate the large asymmetry between transmit power of eNB (≈46dBm) and HeNB(≈20dBm). The algorithm proposed in [10] is depicted in Fig.5. For inbound mobility HO to the femtocell is possible,

$$\text{If, } \bar{s}_f[k] > s_{f,th} \text{ and } s_{pro}^{\alpha}[k] > \bar{s}_m[k] + HMM \qquad (4)$$

$$\text{or} \quad \text{if, } \bar{s}_f[k] < s_{f,th} \text{ and } \bar{s}_f[k] > \bar{s}_m[k] + HMM \qquad (5)$$

On the other hand, for connecting to macrocell from femtocell is made,

$$\bar{s}_f[k] < s_{f,th} \text{ and } \bar{s}_f[k] + HMM < \bar{s}_m[k] \qquad (6)$$

$$\text{or} \quad \text{if, } \bar{s}_f[k] > s_{f,th} \text{ and } s_{pro}^{\alpha}[k] < \bar{s}_m[k] + HMM \qquad (7)$$

The advantages of this algorithm are, it considers the asymmetry in transmit powers between eNB and HeNB and it also includes optimization parameter for the trade-off between HO probability and number of HO failure. Nevertheless, UE speed, bandwidth availability, user subscription and interference were not considered in this single macro-femto model

### 4.1.2 Received Signal Strength and Path Loss Based HO Algorithm:

The proposed HO decision algorithm in [11] considers path loss along with RSS for inbound mobility to femtocells. Similar to the proposals in [10], this path-loss based algorithm also considers exponential window function *w(k)* on the RSRP measurements to compensate the asymmetry level between macro-femto RS transmission powers. Handover to femtocell from macrocell is possible if a) the filtered RSRP measurement of the femtocell exceeds over a minimum threshold, denoted by $RSRP_{th,f}$, b) the filtered RSRP status of the femtocell exceeds over the filtered RSRP status of the macrocell plus the HHM, and c) the observed path-loss between user and FAP is less than the path-loss between UE and the macrocell. The HO algorithm flowchart is illustrated in Fig.6 [11].

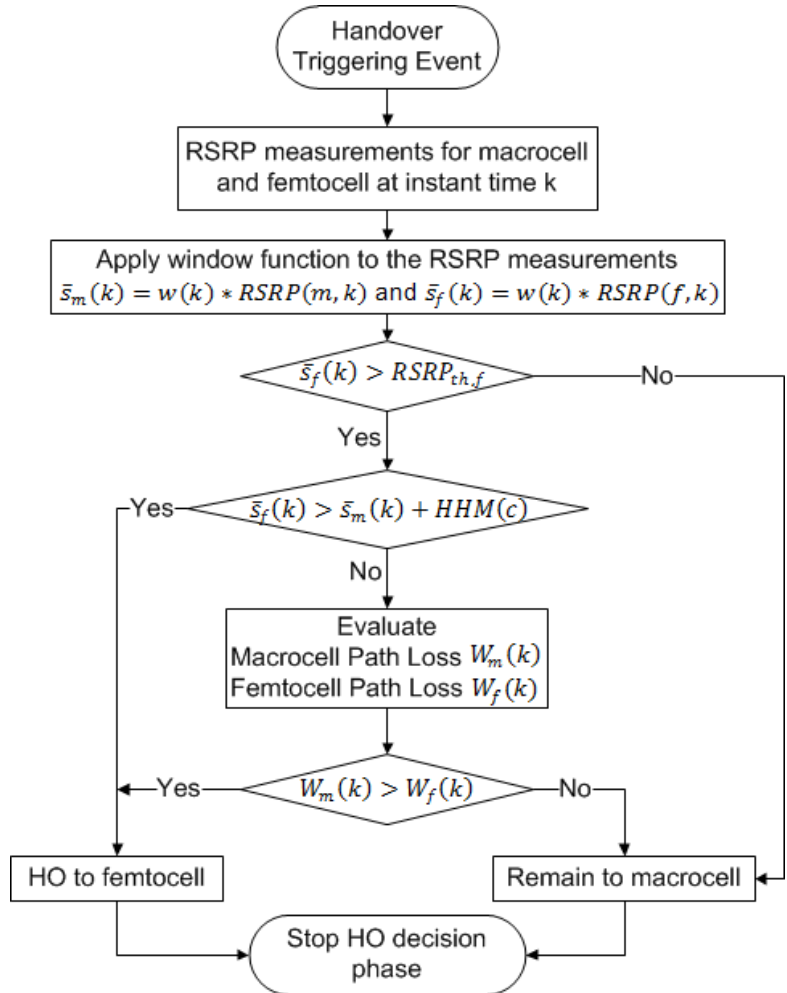

Fig. 6: HO Algorithm based on RSS and Path loss [11]

The main feature of this algorithm is that, it considers actual path-loss between UE and the target cell. However, the algorithm considers single macro-femto model which may not be realistic. On the other hand, path-loss is prone to fast variation which will in turn influence ping-pong effect while HO.

## 4.2 Speed based HO Algorithm:

The main decision criterion for this type of algorithm is speed. In [12], authors proposed HO algorithm conceiving two decision parameters, speed and traffic type. Based on speed either proactive or reactive HO decisions are performed. Proactive HO is one where HO takes place before RSS of the serving cell reaches a pre network defined hysteresis margin. In this type of HO strategy a residual time prior HO execution is estimated. To minimize the HO delay and packet loss for real time traffic is the purpose of pro-active handover. In reactive handover, HO is execution is initiated when minimum required RSS is reached. The purpose of reactive HO is to reduce ping-pong effect. Fig.7 illustrates the operation of multiple macrocell-femtocell scenarios. In the below figure, when the UE speed is higher than 10kmph, there will be no HO from macro to femtocell. When UE has the speed between 5 to 10kmph then this algorithm performs prediction model using Markov-Chain [13] to predict the direction of the user using current position and speed. If the UE moves towards the femtocell, then the proposed model performs either proactive HO if the traffic is real time or reactive HO if the traffic is non-real time. Same approach is made for the users below 5kmph without mobility prediction. This proposed algorithm expected to reduce the HO probability for the users with medium speed (5kmph<speed<10kmph) and better QoS for the real-time traffic users.

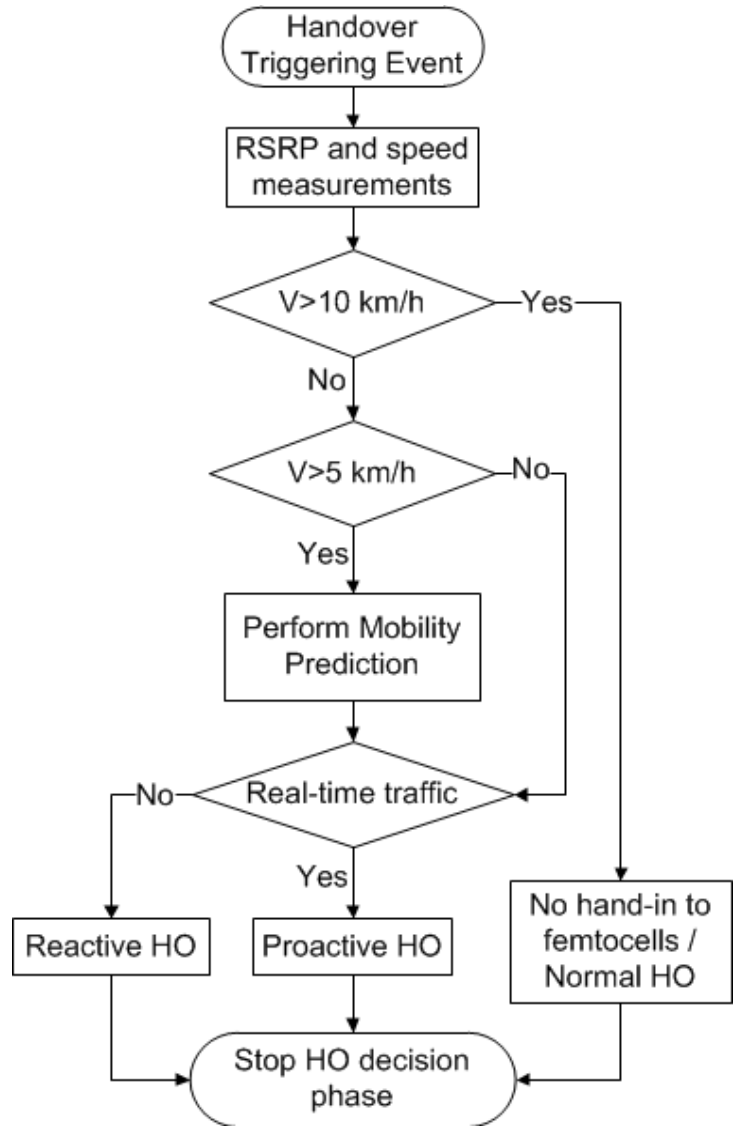

Fig. 7: Speed based HO algorithm [12]

## 5. A Novel Speed Based Algorithm

Femtocells are connected to network through backhaul broadband connection. Due to the absence of X2 interface, the HO decision and execution phase take more time than the conventional macro-cellular handover. Moreover, FAPs have less computational capability. But HO decision algorithms based on energy efficiency [16, 17] or cost function [15] include complex algorithms for FAP to manage and it will also delay the HO decision procedure. Keep these factors in mind; we tried to build up a simple speed based HO algorithm that also includes the other handover parameters i.e., a) access control b) bandwidth satisfaction c) received signal strength and d) traffic type. Our model shown in Fig. 8 is inspired by speed based HO algorithm proposed in [12] .Unlike the model [12], we didn't consider Markov-Chain prediction model since MME knows the speed and location of user [18], which makes the HO decision simple and faster.

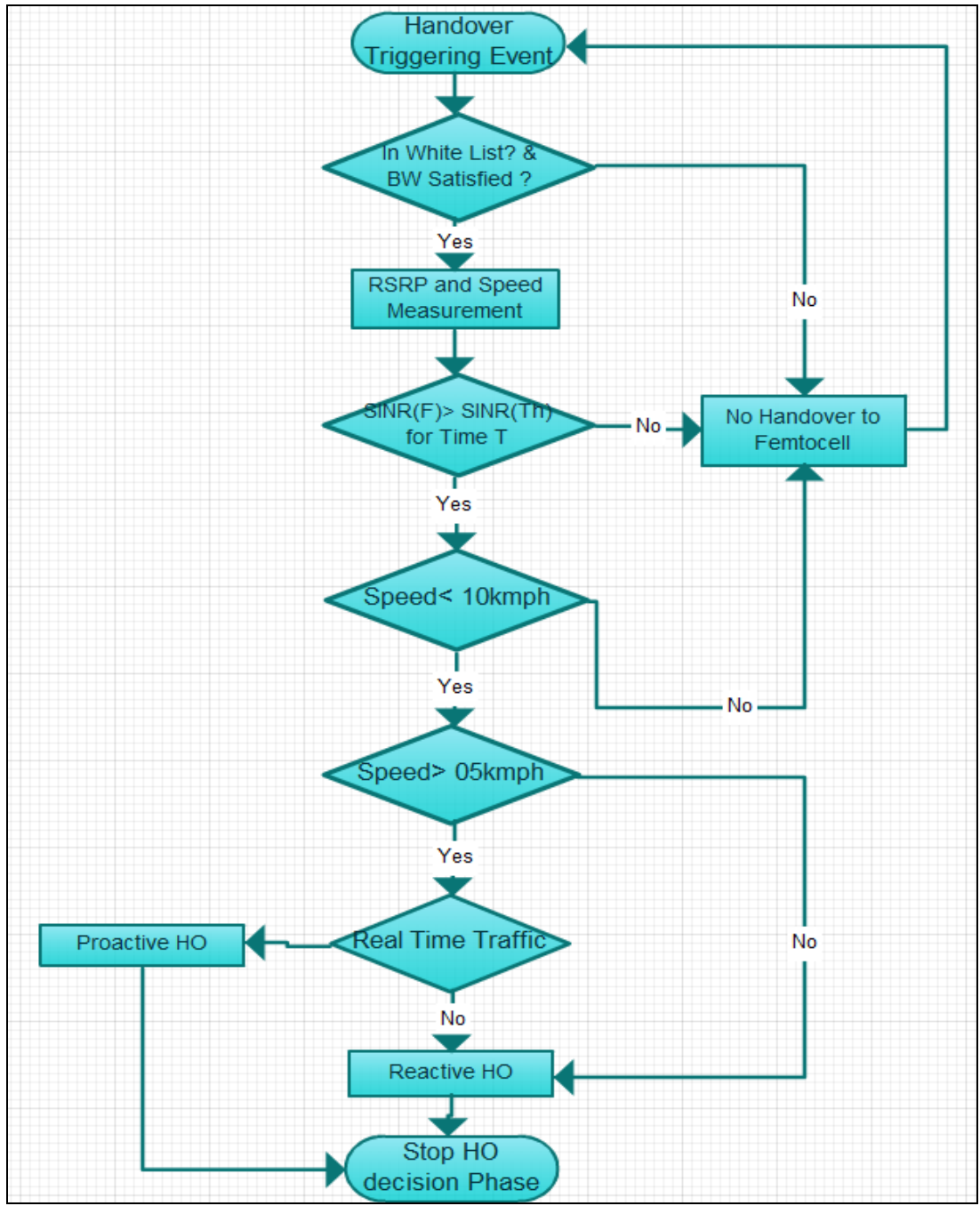

Fig. 8: Proposed Speed Based Handover Model

We used LTE-Sim (an event driven object oriented simulator written in C++ [19]) for simulating inbound Handover. Different packet scheduler (PF, EXP-PF, MLWDF, FLS) are available in LTE-Sim module eNB to perform data flow and resource allocation. The performance of the packet schedulers (PF, M-LWDF and EXP/PF) considering all the users are experiencing single flow (50% of the users are having VoIP flow and the rest are having Video flow) modelled with infinite buffer application was measured prior simulating the proposed algorithm. Fig. 9(a), 9(b) and 9(c) shows the performance of three different packet schedulers in terms of Throughput, Packet Loss Ratio (PLR) and Spectral efficiency considering users are moving with the speed of 3Kmph (Pedestrian Speed) and 120Kmph (Highway Speed). While simulating Random walk model [20] was considered. All the results show in all cases in LTE-Sim, MLWDF has the better performance. The simulation parameters are mentioned in Table-1.

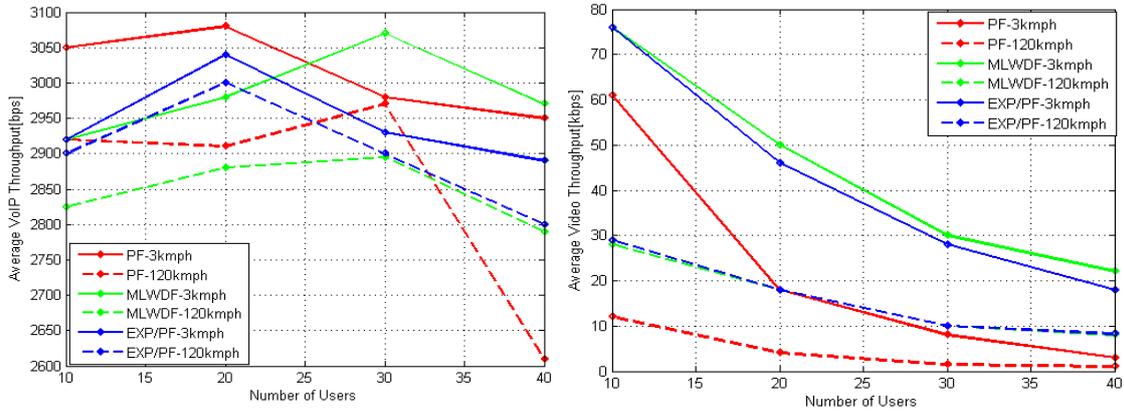

Throughput (VoIP)          Throughput (Video)

Fig. 9 (a)   Average Throughput of **VoIP** and **Video Flow** with different schedulers at different speed [21]

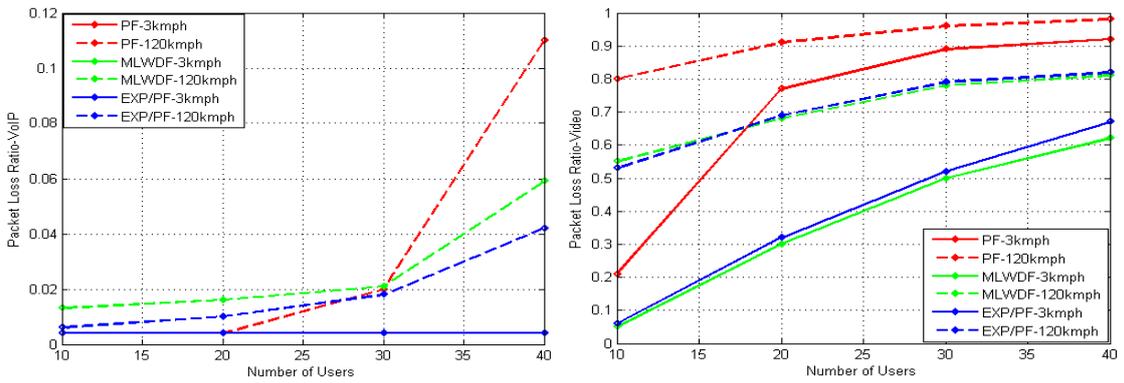

PLR (VoIP)          PLR (Video)

Fig. 9 (b)   Packet Loss Ratio (PLR) of **VoIP** and **Video Flow** with different schedulers at different speed [21]

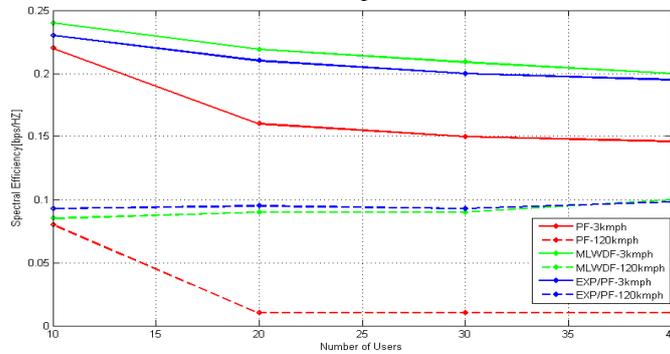

Fig. 9(c) Spectral Efficiency of different schedulers in LTE-SIM [21]

For the simplication of the HO algorithm, PF was considered in our proposed HO algorithm. Fig. 10 shows the comparison between the assignment probabilities of our proposed model to the earlier mentioned models of RSS based HO model [11] and speed based HO model [12]. From Fig. 10 it can be seen that, our proposed HO model is showing better performance when FAP is located near the eNB and the assignment probability to FAP is higher than other two models.

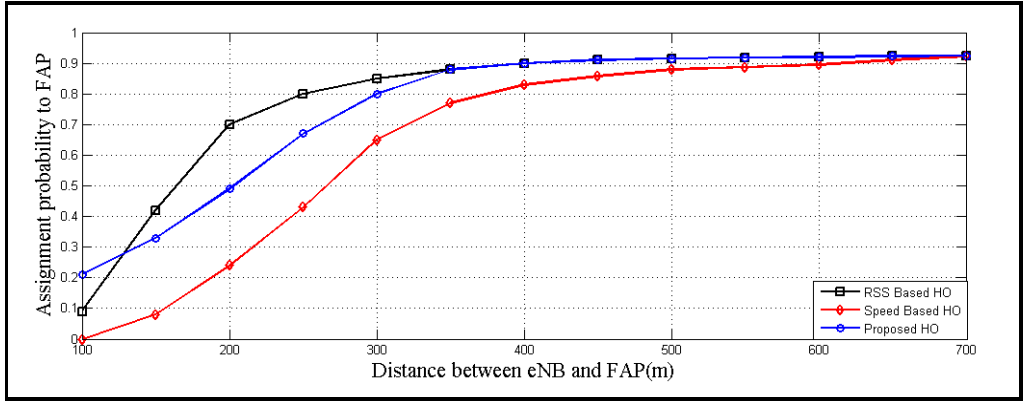

Fig. 10: Assignment Probability to FAP vs. distance between eNB and FAP

Table 1: Simulation Parameters.

| **Total Bandwidth** | 20 MHz |
|---|---|
| eNb power transmission | 43 dBm |
| FAP power transmission | 20 dBm |
| CQI | Full Bandwidth and periodic (2ms) reporting scheme |
| Apartment Size | 100 $m^2$ |
| Building Type | 5*5 Apartment grid |
| Number of FAPs | 1 FAP/Apartment |
| CSG Users | 9 FUE/FAP |
| Scheduler | PF |
| Traffic | VOIP, VIDEO |
| Mobility | Random Walk Model |

In Fig.11 it can also be seen that, in our proposed model the number of handover is less than RSS based or Speed based HO models within 350 meters. But after 350 meters it shows higher handoffs than the RSS based HO model, because at the cell edge the downlink interference experienced because of the presence of MUE near FAP is higher.

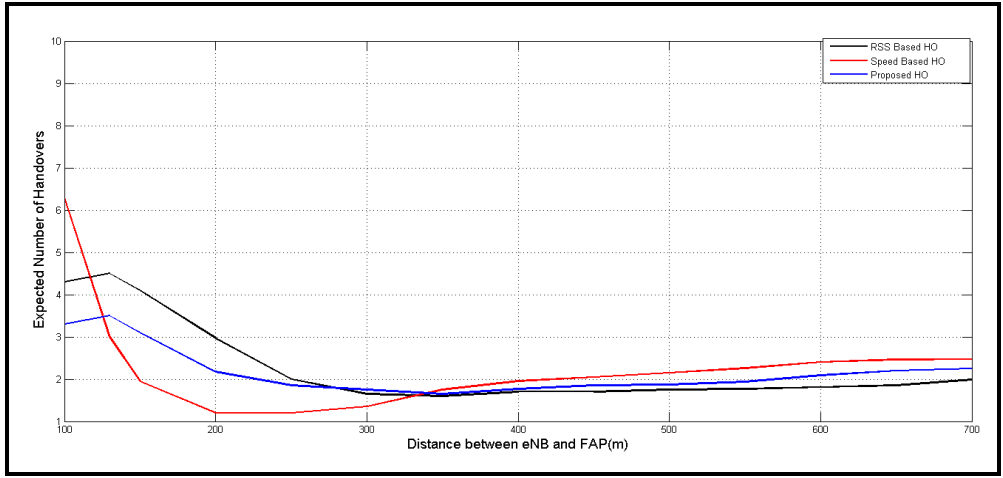

Fig. 11: Number of Expected Handovers vs. Distance between eNB and FAP

Table 2: Features of the Proposed Speed Based HO Algorithm

| Signal Strength | Strengths | Future Modification |
|---|---|---|
| Minimum RSS for HO | √ | |
| Path Loss | √ | |
| Window Function | √ | |
| SINR | √ | |
| **Speed** | | |
| UE Speed | √ | |
| UE mobility Pattern | | X |
| **BW Related** | | |
| Cell Capacity | √ | |
| Cell Load | √ | |
| Number of UE's Camped | | X |
| Cell Type | √ | |
| **Traffic Related** | | |
| Traffic Type | √ | |
| BER | | X |
| **Energy Efficiency Related** | | |
| UE power Class | | X |
| UE battery Class | | X |
| Mean UE transmit Power | | X |
| **Access** | | |
| UE membership Status | √ | |
| UE Application priorities | √ | |
| UE Priorities | | X |

# 6. CONCLUSION

In our paper, we tried to show the mathematical complexity of the renowned approaches for HO techniques and comparisons of the simulation result of the HO algorithm to others. The main aim was to create a speed based algorithm, though our HO decision technique also considers all the aspects of HO i.e., bandwidth, SINR (interference), traffic type, access permission as illustrated in Table-2. The preliminary simulation results show a better result than the signal strength based and speed based HO techniques and showing lower number of HO attempts in the cell centre areas. However, for future work we are analysing to improve our proposed algorithm to be energy efficient and to make it performs better in the cell edge area from other HO algorithms.

**Authors**

**Pantha Ghosal** is a Graduate Research Assistant at Faculty of Engineering and Technology (FEIT), CRIN, University of Technology, Sydney. Prior to this, he completed B.Sc in Electrical and Electronic Engineering from Rajshahi University of Engineering & Technology, Bangladesh in 2007. He is an expert of Telecommunication network design and holds more than 7 years of working experience in 2G/3G and LTE. Throughout his career he was involved in several projects of RF Planning, Designing and Dimensioning.

**Shouman Barua** is a PhD research scholar at the University of Technology, Sydney. He received his BSc in Electrical and Electronic Engineering from Chittagong University of Engineering and Technology, Bangladesh and MSc in Information and Communication Engineering from Technische Universität Darmstadt (Technical University of Darmstadt), Germany in 2006 and 2014 respectively. He holds also more than five years extensive working experience in telecommunication sector in various roles including network planning and operation.

**Ramprasad Subramanian** is an experienced telecom engineer in the field of 2G/3G and LTE/LTE-A. He holds M.S (By research) in Information and Communication from Institute of Remote Sensing, Anna University (India)(2007) and Bachelors of Engineering in Electronics and Communication engineering from Bharathidasan University (2001)(India). He has done many projects in the area of 2G/3G and LTE. He has done many consultative projects across Africs/Americas/Asia etc. He was the recipient of India's best invention award for the year 2004 from Indian Institute of Management Ahmadabad and Government of India. His current research focuses on 4G mobile networks and vehicular Ad hoc networks.

**Shiqi Xing** is currently doing his Bachelor of Telecommunication Engineering at University of Technology, Sydney. He is an experienced programmer and currently undertaking projects in 4G Telecommunication and Robotics. Throughout his academic career he received several scholarships from The People's Republic of China.

**Dr Kumbesan Sandrasegaran** is an Associate Professor at UTS and Centre for Real-Time Information Networks (CRIN). He holds a PhD in Electrical Engineering from McGill University (Canada)(1994), a Master of Science Degree in Telecommunication Engineering from Essex University (1988) and a Bachelor of Science (Honours) Degree in Electrical Engineering (First Class) (1985).  His current research work focuses on two main areas (a) radio resource management in mobile networks, (b) engineering of remote monitoring systems for novel applications with industry through the use of embedded systems, sensors and communications systems. He has published over 100 refereed publications and 20 consultancy reports spanning telecommunication and computing systems.